\documentclass{webofc}
\usepackage[varg]{txfonts}   % Web of Conferences font

\woctitle{?????????}
\begin{document}
\title{Triggered fragmentation in gravitationally unstable discs: forming fragments at small radii}

\author{Farzana Meru\inst{1,2,3}\fnsep\thanks{\email{farzana.meru@phys.ethz.ch}} 
}

\institute{
Institut f\"ur Astronomie, ETH Z\"urich, Wolfgang-Pauli-Strasse 27, 8093 Z\"urich, Switzerland
\and
Institut f\"ur Astronomie und Astrophysik, Universit\"at T\"ubingen, Auf der Morgenstelle 10, 72076 T\"ubingen, Germany
\and
School of Physics, University of Exeter, Stocker Road, Exeter, EX4 4QL, UK
         }

\abstract{
We carry out three dimensional radiation hydrodynamical simulations of gravitationally unstable discs using to explore the movement of mass in a disc following its fragmentation.  Compared to a more quiescent state before it fragments, the radial velocity of the gas increases by up to a factor of $\approx 2-3$ after fragmentation.  While the mass movement occurs both inwards and outwards, the inwards motion can cause the inner spirals to be sufficiently dense that they may become unstable and potentially fragment.  Consequently, the dynamical behaviour of fragmented discs may cause subsequent fragmentation at smaller radii \emph{after} an initial fragment has formed in the outer disc.
}

\maketitle

\section{Introduction}
\label{intro}

Gravitationally unstable discs may collapse and fragment to form bound objects if: i) they become unstable such that the Toomre stability parameter \cite{Toomre_stability1964}, $Q = \frac{c_{\rm s} \kappa}{\pi \Sigma G} \lesssim 1$, where $c_{\rm s}$ is the sound speed in the disc, $\kappa_{\rm ep}$ is the epicyclic frequency, which for Keplerian discs is $\approx \Omega$ (the angular frequency), $\Sigma$ is the surface mass density and $G$ is the gravitational constant, and ii) the cooling timescale is faster than a critical rate, $t_{\rm cool} < \beta_{\rm crit}/\Omega$ \cite{Gammie_betacool}.  The value of $\beta_{\rm crit}$ has been the subject of much recent research, the latest of which suggests that $\beta_{\rm crit} > 20$ \cite{Meru_Bate_convergence}.  On large scales AGN discs may fragment to form stars \citep[e.g.][]{Nayakshin_AGN_fragmentation} while for smaller scale discs, brown dwarfs \cite[e.g.][]{Stamatellos_BD} or even planets \cite{GI_Cameron} may form.

Upon formation, the fragment-disc interactions (angular momentum exchanges and mass accretion onto the fragment) may cause fragment migration \citep[e.g.][]{Baruteau_GI_migration,Michael_GI_migration} or even disc disruption \cite[e.g.][]{Stamatellos_BD}.  However, while the longer term fragment evolution has been considered in some detail, how the disc mass moves in response to the presence of the fragment immediately after its formation is not so well studied.  In the context of planets, the existence of a planet in a gravitationally unstable disc can trigger the collapse of the disc to form fragments both interior and exterior to it \cite{Armitage_Hansen_trigger}.  However, this study used an isothermal equation of state (equivalent to fast cooling) and was thus favourable for fragmentation.

\section{Simulations}

\begin{table}
\centering
\caption{Table showing the simulation parameters}
\label{tab:sim}
\begin{tabular}{clllcll}
\hline
Simulation & $M_{\star} [M_{\odot}]$ & $M_{\rm disc} [M_{\odot}]$ & $L_{\star} [L_{\odot}]$ & Opacity fraction & Initial $\Sigma$ profile & Initial $T$ profile \\\hline
1 & 1.5 & 1.2 & 4.3 & 1/3 & $\Sigma \propto R^{-3/2}$ & $T \propto R^{-1/2}$  \\
2 & 0.8 & 0.8 & 1.7 & 1/10 & $\Sigma \propto R^{-3/2}$ & $T \propto R^{-1/2}$ \\\hline
\end{tabular}
\end{table}

We carry out three-dimensional Smoothed Particle Hydrodynamics simulations with radiative transfer of self-gravitating discs to understand the mass movement in the discs before and after fragmentation.  We self-consistently model the formation of the first and any potential subsequent fragments.  The disc surface mass density and temperature are chosen such that the Toomre profile decreases with radius so that the outer disc has the opportunity to fragment while the inner disc has $Q \gtrsim 1$ and is thus not expected to fragment.  We carry out two main simulations (Table~\ref{tab:sim}).  The initial temperature (and the disc's upper boundary temperature) is set by the central star's luminosity (using $L_{\star} = 4 \pi R^2 \sigma T^4$, where $\sigma$ is the Stefan-Boltzmann constant) to take into account stellar irradiation.  We model the discs using Rosseland mean opacities that are a fraction of the interstellar values (see Table~\ref{tab:sim}) as done previously \cite{Meru_Bate_opacity}.  To follow the subsequent disc evolution, any fragments are turned into sink particles.

Furthermore, we simulate a non-fragmenting gravitationally unstable disc to compare the mass movement in fragmenting discs with \emph{quiescent} ones.  This disc is the same as Simulation 2 but modelled using an opacity fraction of 1/3 (thus decreasing the cooling efficiency to inhibit fragmentation).

\section{Results}

\begin{figure*}
\centering
\includegraphics[width=1.0\columnwidth,clip]{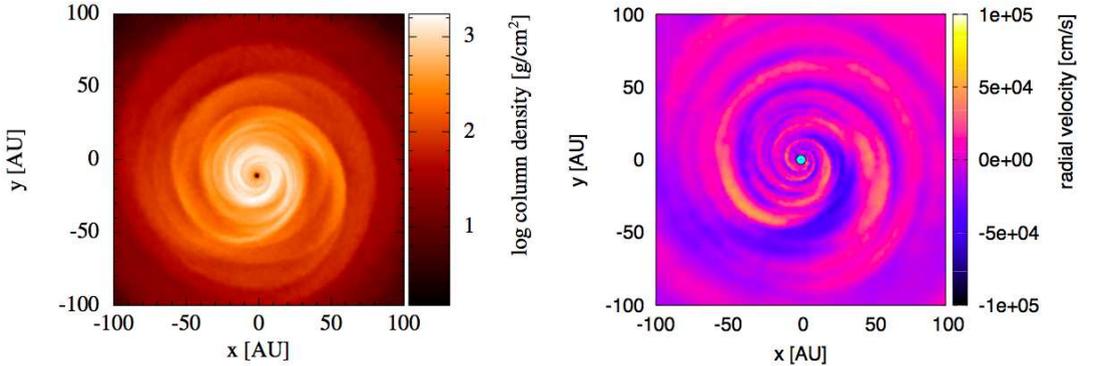}
\caption{Surface mass density (left panel) and radial velocity (right panel) rendered images of a non-fragmenting \emph{quiescent} self-gravitating disc.}
\label{fig:no_frag}     
\end{figure*}

Figure~\ref{fig:no_frag} shows the surface mass density and radial velocity rendered images of the non-fragmenting disc.  In such \emph{quiescent} discs the magnitude of the radial velocity can be up to $\approx 5 \times 10^4$ cm/s.

\begin{figure*}
\centering
\includegraphics[width=0.9\columnwidth,clip]{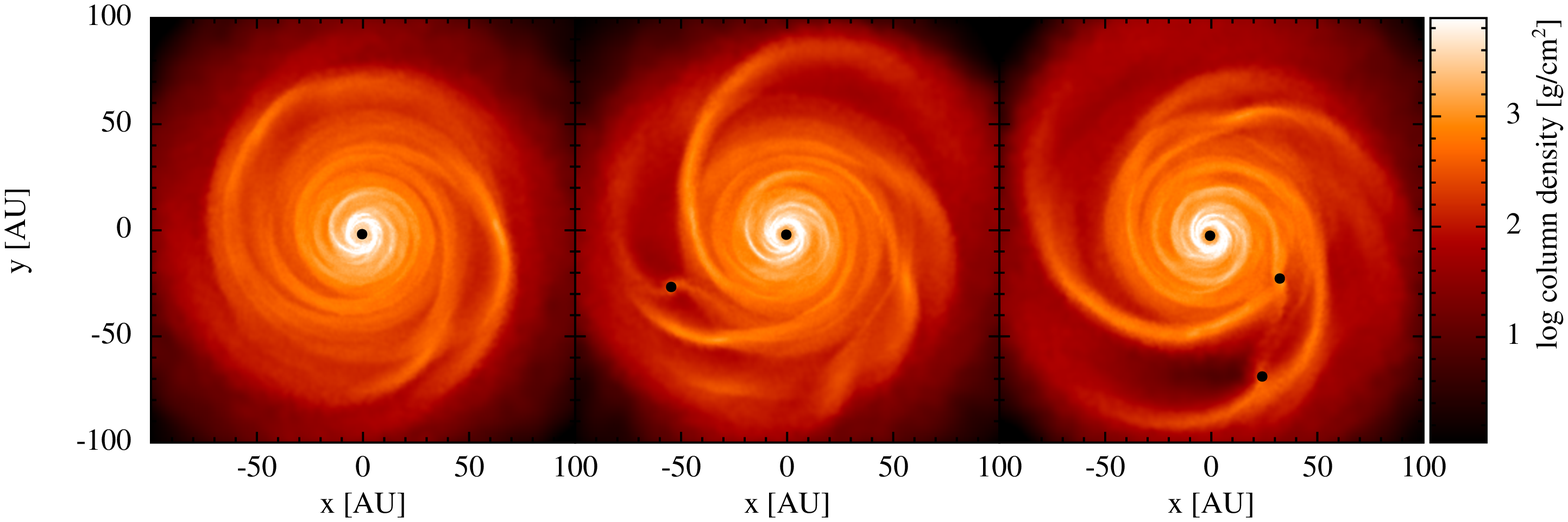}\\
\vspace*{-0.8cm}
\hspace*{0.46cm}
\includegraphics[width=0.350\columnwidth,clip,angle=-90]{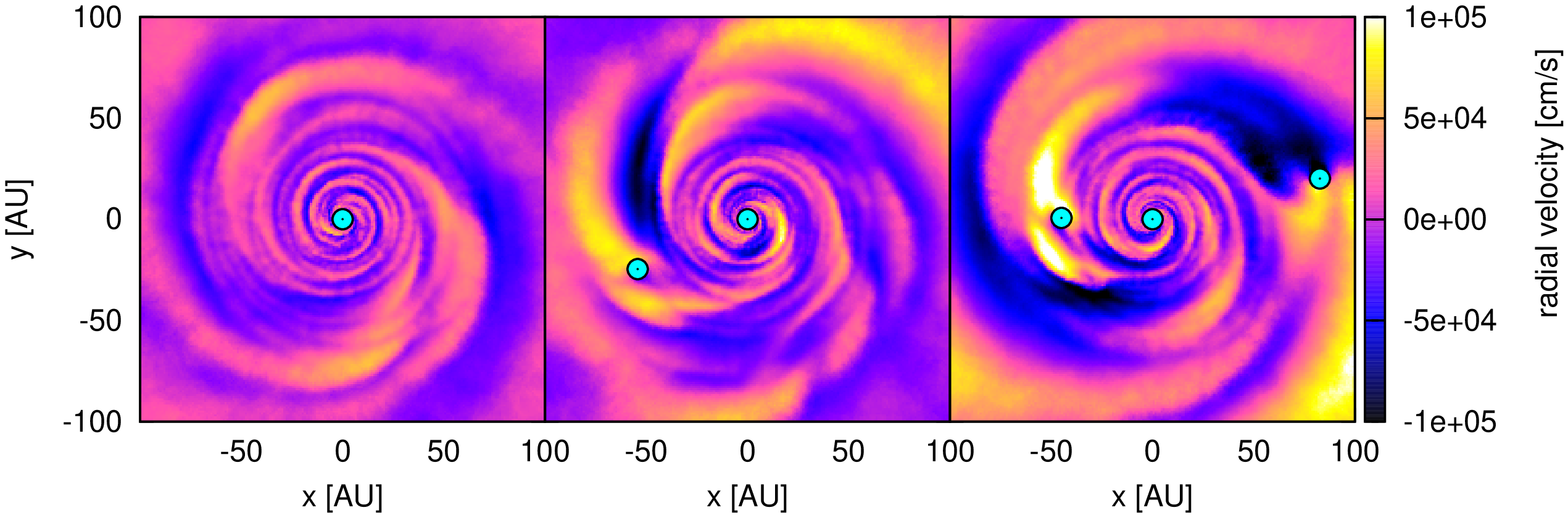} 
\caption{Surface mass density (top panel) and radial velocity (bottom panel) rendered images of the self-gravitating disc in Simulation 1 before the first fragment forms (left panel) and after the formation of the first (middle panel) and second fragments (right panel).  The mass movement increases after the fragments form.  Note that the two images in the right panel are at slightly different times.}
\label{fig:seq_frag} 
\end{figure*}

\begin{figure*}
\centering
\includegraphics[width=0.5\columnwidth,clip]{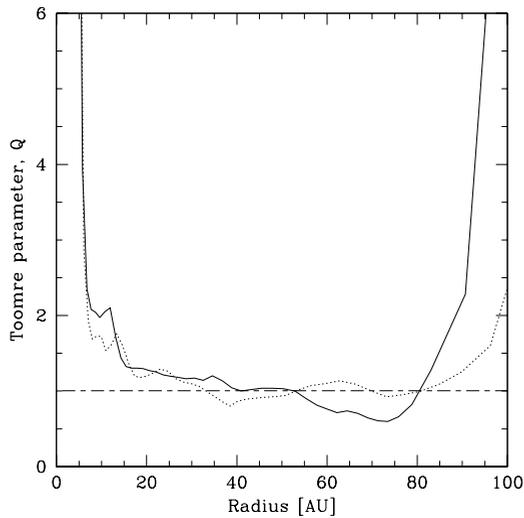}
\caption{Azimuthally averaged Toomre stability profiles of the self-gravitating disc in Simulation 1 shortly before the first (solid line) and second (dotted line) fragments form.}
\label{fig:toomre}       
\end{figure*}

\begin{figure*}
\centering
\includegraphics[width=0.23\columnwidth,clip,angle=-90]{Fig4a.ps}
\includegraphics[width=0.24\columnwidth,clip,angle=-90]{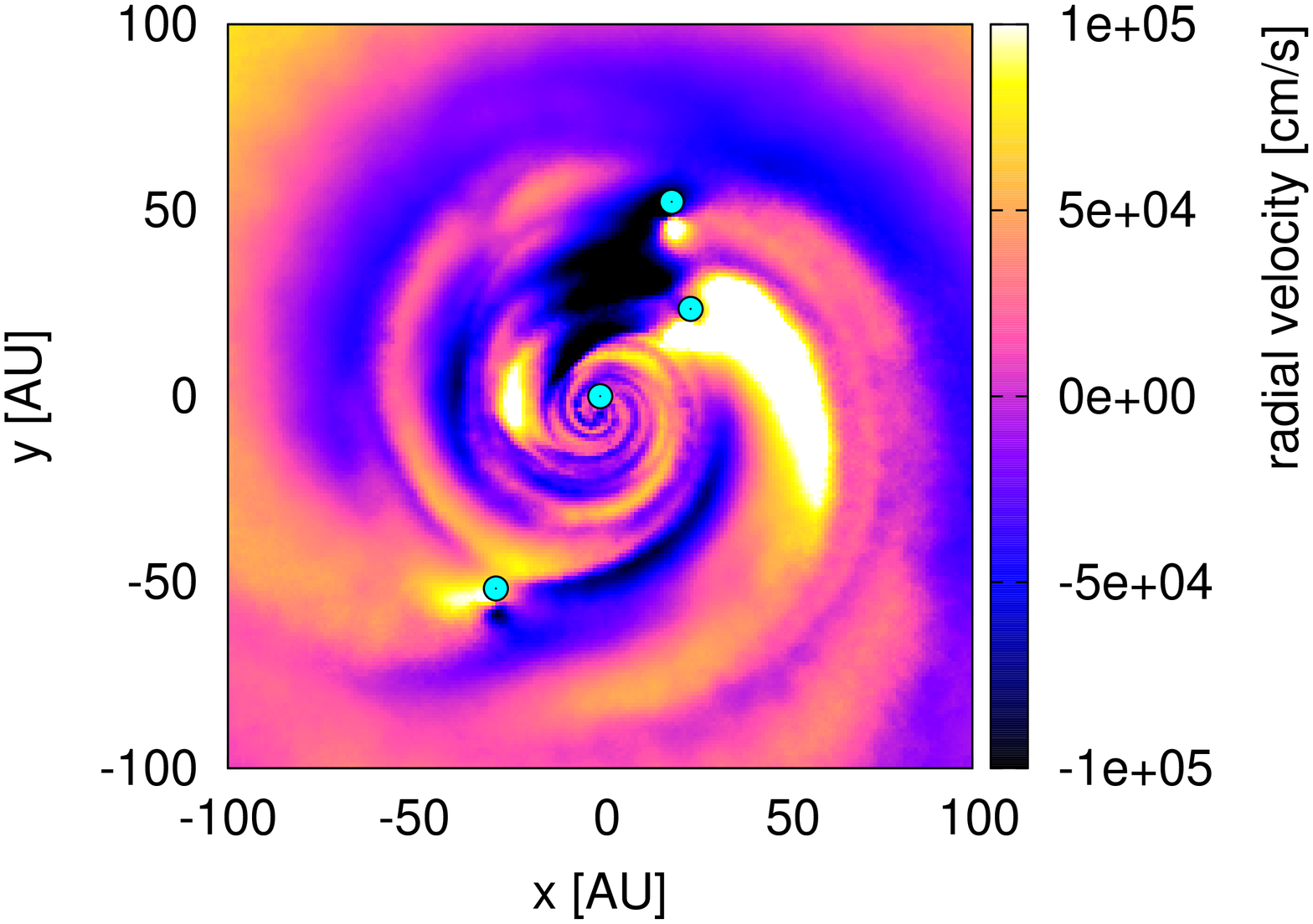}
\includegraphics[width=0.23\columnwidth,clip,angle=-90]{Fig4c.ps}
\caption{Surface mass density and radial velocity rendered images of the self-gravitating disc in Simulation 2 at three evolutionary stages (chronological from left to right).}
\label{fig:sim2}       
\end{figure*}

Figure~\ref{fig:seq_frag} shows the surface mass density (top panel) and radial velocity (bottom panel) rendered images of the disc in Simulation 1 at various evolutionary stages.  Before fragmentation occurs (left panel) the disc is reasonably quiescent, similar to the non-fragmenting disc in Figure~\ref{fig:no_frag}.  However once the first fragment forms at $\approx 60$~AU, its presence causes the radial velocity of the gas in regions of the disc that it has interacted with to increase, both inwards and outwards by a factor of $\approx 2-3$ (middle panel).  The mass movement in the fragmented disc is clearly not quiescent as in the non-fragmenting gravitationally unstable disc (cf Figure~\ref{fig:no_frag} right panel) or in the disc before it fragments.  As a result of the inwards motion, an inner spiral becomes more dense and fragments at $\approx 39$~AU (Figure~\ref{fig:seq_frag}, top right panel).  Figure~\ref{fig:toomre} shows the Toomre stability profile just before the formation of each of the fragments.  Prior to the first fragment forming, a large part of the outer disc is gravitationally unstable (with $Q < 1$) but at $R \lesssim 50$~AU the disc has $Q \gtrsim 1$, and thus is not so susceptible to fragmentation.  However, the mass movement that occurs as a result of the presence of the first fragment causes the inner region of the disc to become dense enough that the Toomre profile decreases, allowing the disc to be pushed into a state of instability, resulting in fragmentation.  Once the second fragment forms, the disc becomes even more dynamic leading to additional regions where mass is moving inwards or outwards at a high rate (Figure~\ref{fig:seq_frag}, bottom right panel).  We note that the inward movement causes more spirals at smaller radii to become increasingly dense, triggering them to fragment.

In Simulation 2 three fragments form out of one spiral arm (Figure~\ref{fig:sim2}, left panel), the presence of which causes much dynamical movement both inwards and outwards (Figure~\ref{fig:sim2}, middle panel).  The inwards movement causes an inner spiral to become more dense and fragment (Figure~\ref{fig:sim2}, right panel).

\section{In the context of protoplanetary discs}

Triggered fragmentation may have important implications for giant planet formation theories: while a disc may only initially fragment in the outer regions \citep{Rafikov_SI,Clarke2009_analytical,Boley_CA_and_GI}, planets may well form in the inner parts via gravitational instability \emph{provided} fragmentation has already occurred in the outer regions first.  Since core accretion is thought to occur at small radii (up to $\approx 5-10$~AU) while gravitational instability is thought to occur at larger radii ($\gtrsim 50$~AU) \emph{subsequent planet formation by gravitational instability} may well be a mechanism that may operate in the radial range in between, where no single in situ formation method is currently thought to dominate.

\section{Summary}
We carry out 3D radiative transfer simulations of gravitationally unstable discs to explore how the disc mass moves following its fragmentation.  We model the formation of the first fragment and subsequent disc evolution self-consistently.  The fragment's presence causes the magnitude of the radial velocities to increase by a factor of $\approx 2-3$ resulting in a more dynamic disc than before it formed.  If an inner disc region is quite close to marginal stability, such mass movement can cause it to be pushed into a state of instability where further fragmentation may then occur.  This has potential implications for planet formation at the radii where no one mechanism can adequately describe their in situ formation.

% BibTeX or Biber users please use (the style is already called in the class, ensure that the "woc.bst" style is in your local directory)
\bibliography{references}
\bibliographystyle{woc}

\end{document}